\documentclass[pra,twocolumn,floatfix,nofootinbib,showpacs]{revtex4}

\usepackage{amsmath,amssymb,xspace,graphicx,units,hyperref,verbatim,upgreek}

\renewcommand{\vec}[1]{\boldsymbol{#1}}
\newcommand{\rhos}{\ensuremath{n_{\text{s}}}\xspace}
\renewcommand{\d}{\ensuremath{\mathrm{d}}}

\newcommand{\D}{\ensuremath{\mathcal{D}}}

\newcommand{\V}{\ensuremath{\mathcal{V}}}

\newcommand{\N}{\ensuremath{\mathcal{N}}\xspace}

\newcommand{\ie}{\emph{i.e.}\xspace}
\newcommand{\eg}{\emph{e.g.}\xspace}

\newcommand{\zetanb}[1]{\ensuremath{\zeta_{\{\vec{k}\}}^{(#1)}\xspace}}
\renewcommand{\Im}{\ensuremath{\text{Im}}}
\newcommand{\hc}{\ensuremath{\text{h.c.}}}
\newcommand{\normorder}[1]{\ensuremath{:\!#1\!:}\xspace}
\newcommand{\abs}[1]{\left|#1\right|}
\newcommand{\etal}{\emph{et al.}\xspace}
\newcommand{\Tr}{\ensuremath{\text{Tr}}}
\newcommand{\kB}{\ensuremath{k_{\text{B}}}}
\newcommand{\micron}{\ensuremath{\upmu\mathrm{m}}}

\begin{document}
\title{Full counting statistics of the interference contrast from
independent\\Bose-Einstein condensates}
\author{Steffen Patrick Rath and Wilhelm Zwerger}
\affiliation{Technische Universit{\"a}t M{\"u}nchen, Physik Department,
James-Franck-Stra{\ss}e, 85748 Garching, Germany}
\begin{abstract}
We show that the visibility in interference experiments with Bose-Einstein
condensates is directly related to the condensate fraction. The probability
distribution of the contrast over many runs of an interference experiment thus
gives the full counting statistics of the condensed atom number. For
two-dimensional Bose gases, we discuss the universal behavior of the
probability distribution in the superfluid regime and provide analytical
expressions for the distributions for both homogeneous and harmonically trapped
samples. They are non-Gaussian and unimodal with a variance that is directly
related to the superfluid density.  In general, the visibility is a
self-averaging observable only in the presence of long range phase coherence.
Close to the transition temperature, the visibility distribution reflects the
universal order parameter distribution in the vicinity of the critical point. 
\end{abstract}

\maketitle

\section{Introduction}
\label{sec:intro}

Interference experiments constitute an invaluable tool for the characterization
of the coherence properties of ultracold gases
\cite{andr1997,hadz2006,hoff2007a}. These properties are particularly
intriguing in the case of ultracold one- or two-dimensional Bose gases
\cite{petr2000a,petr2000,bloc2008}. Due to strong
phase fluctuations, both the 1d gas at zero temperature and the 2d gas at
finite temperature exhibit only quasi-long-range order, \ie, the one-body
density matrix which measures long range phase coherence decays as a power law
instead of converging to a finite constant for long distances. Due to the
quantum nature of the interfering matter fields, the measured visibility is a
random variable that differs from one experimental run to the other. For the
case of one-dimensional Bose gases at zero temperature, the probability
distribution of the interference contrast has been calculated analytically for
arbitrary strong interactions, using a mapping to the exactly solvable
boundary sine-Gordon theory \cite{grit2006}. In the weak coupling limit, it
turns out to be a Gumbel distribution \cite{imam2006,imam2008}.  Numerical data
calculated within the same theoretical framework, but at finite temperature,
are in good agreement with experiment \cite{hoff2008}. 

Here we reconsider the issue of the statistics of the interference contrast for
Bose-Einstein condensates  in arbitrary dimension, discussing in particular the
2d case and the connection between long range order and self-averaging.  We
show that the probability distribution of the interference contrast  is
identical to that of the condensate fraction in the limit of a large
integration volume in the absorption images. The resulting distributions
therefore provide the precise counting statistics for the number of condensed
atoms \cite{naza2009}.  The statistics of the condensate number has a
universal form in two limiting cases: at the critical point as a consequence of
finite size scaling \cite{fish1972,brez1985} and at temperatures far
below the critical temperature. We discuss both cases and give analytical
expressions for the distribution for the latter. Specifically, for 2d Bose
gases, the distribution at temperatures $T$ far below the critical temperature
of the Berezinskii-Kosterlitz-Thouless (BKT) transition
\cite{bere1971,kost1973}, is controlled by the dimensionless parameter
$\eta(T)=1/\rhos\lambda_T^2$, where $\rhos$ is the superfluid density, and
$\lambda_T=\sqrt{2\pi\hbar^2/m\kB T}$ the thermal wavelength.  The
fluctuations around the average visibility are determined by the superfluid
density, a quantity that is rather difficult to measure by other means
\cite{coop2010}.  For a homogeneous square sample, we show that the probability
distribution of the interference contrast is close to a convolution of two
Gumbel distributions, similar to but different from the Gumbel distribution
that is obtained for the weakly interacting limit of a 1d Bose gas at zero
temperature \cite{imam2006,imam2008}. Non-Gaussian distributions are also found
in harmonically trapped and strongly anisotropic 2d gases, in qualitative
agreement with preliminary data taken at the Ecole normale sup{\'e}rieure (ENS)
in Paris \cite{hadzunpu}.  The principal
focus of this article is on the physics of 2d Bose gases, but we will highlight
the differences and similarities to the 3d case as well as the 1d case at
vanishing temperature.

This article is organized as follows. In section \ref{sec:intstats}, we
introduce the physical system and discuss the connection between the measured
distribution of the interference contrast and that of the condensed fraction.
In section \ref{sec:gaussian} we use a functional integral description for
explicit analytical or numerical calculations of the visibility distribution in
the regime where phase fluctuations are dominant.  In particular, we discuss
the general form of the probability distribution in terms of its cumulants and
the issue of self-averaging.  Section \ref{sec:analyt} is devoted to the
explicit analytical calculation of the probability distribution in the 2d case
far below the BKT transition temperature for both a homogeneous system and a
harmonically trapped sample.  In section \ref{sec:critpoint}, we discuss the
scaling behavior of the probability distribution of the interference contrast
at the critical point, where the average visibility vanishes.  The 
one dimensional case and the anomalous fluctuations  of the condensate fraction
in three dimensions are discussed in the appendices.

\begin{figure}[htbp]
	\centering
	\includegraphics[width=.48\textwidth]{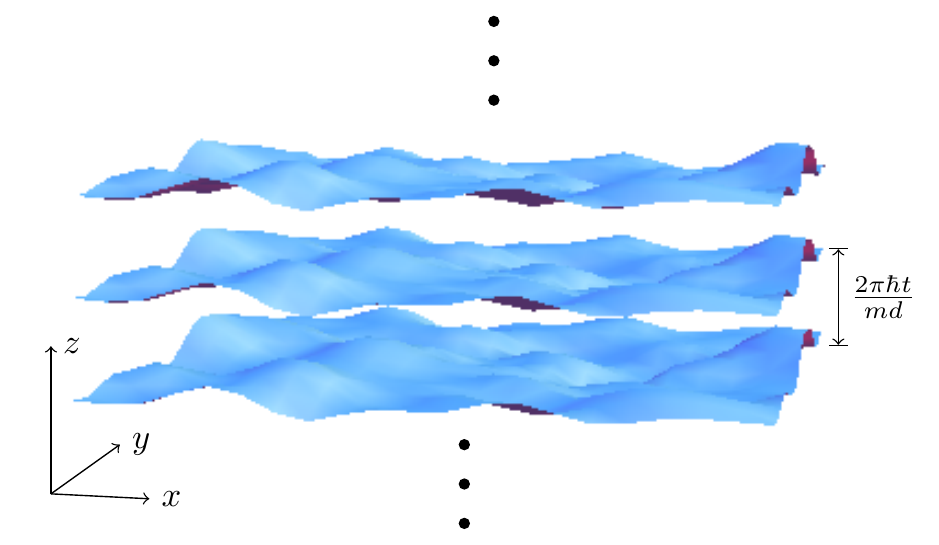}
	\caption{(Color online) Illustration of the experimental situation in
	the 2d case.  After a time $t$ of rapid expansion, two identical Bose
	gases overlap and interfere. The interference pattern is characterized
	by its iso-phase surfaces. Three surfaces having all the same phase are
	shown for illustration, the pattern itself continues all over the
	vertical extension of the sample, as suggested by the continuation dots
	above and below the shown surfaces. The sample is imaged using an
	absorption beam in the $y$ direction.}
	\label{fig:setup}
\end{figure}

\section{Interference statistics}
\label{sec:intstats}

Typical interference experiments with 2d Bose gases 
(\eg, \cite{hadz2006,krug2007}) start out by preparing a
pair of such gases confined to the lowest transverse mode in the $z$ direction
and 
separated by a distance $d$ in the $z$ direction. The atoms are then released
from the trap and imaged after an adjustable free expansion time using
absorption imaging (time of flight measurement). When the trapping potentials
are cut off, 
the gases rapidly expand in the $z$ direction while the density
distribution as a function of $x$ and $y$ can approximately be regarded as
constant. Within this initial expansion period, there is no transformation of
phase fluctuations into density fluctuations, which only sets in at a later
stage \cite{imam2009}.
Due to the rapid expansion along $z$, the gases completely overlap after a
time $t$ of the order of \unit[20]{ms} for typical traps.
The difference between their phases then results in
a spatially modulated interference pattern (see Fig.~\ref{fig:setup}).

The operator for the atomic density after time of flight at an observation
point $\vec{x}$ can be written as
\cite{polk2006}
\begin{equation}
	\hat{n}(\vec{x})=n_0(z)\left[ 
	\hat{n}_1(\vec{r})+\hat{n}_2(\vec{r})+\hat{A}(\vec{r})e^{iQz}+\hc
	\right]\ .
	\label{eq:totdens}
\end{equation}
Here, $n_0(z)$ is an envelope function (\ie, a normalized function of $z$ which has
a negligible Fourier component at the wave number $Q$ of the expansion),
$\hat{n}_{1,2}$ are the in situ (\ie, before time of flight) density
operators of the individual gases, $Q=md/\hbar t$ the wave vector associated with
the ballistic expansion and $\hat{A}(\vec{r})=\hat{\psi}_1^\dagger(\vec{r})\hat{\psi}_2(\vec{r})$ 
the operator that determines the local interference
amplitude. Here and in the following, we use $\vec{r}$ to denote a point in the
trap before time of flight, and $\vec{x}$ for observation points after time of
flight \cite{bloc2008}. 

In a given experimental run, the measured density distribution $n(\vec{x})$
corresponds to an eigenstate of the hermitian operator $\hat{n}(\vec{x})$ and
$\hat{A}(\vec{r})$ can be replaced by a complex number $A=\V(\vec{r})
e^{i\phi(\vec{r})}$. The resulting density can then be written in the
standard form $n_0(\vec{x})(1+\V(\vec{r})\cos\{Q[z-z_0(\vec{r})]\})$ of an
interference pattern with the local visibility $\V(\vec{r})=|A|$ and a
spatially varying shift $z_0(\vec{r})=-\phi(\vec{r})/Q$. When the
interference pattern is integrated over a finite volume, spatial variations of
both $\V(\vec{r})$ (caused by density fluctuations) and $z_0(\vec{r})$ (caused
by phase fluctuations) lead to a reduction of the integrated visibility $\V$.
The height function $z_0$ defines a surface in real space, the
iso-phase surface. In one or two dimensions, there is one unique iso-phase
surface which is repeated over the entire extension of the sample (cf.
Fig.~\ref{fig:setup}). In three dimensions, $z_0$ depends on all three spatial
coordinates and the shape of the iso-phase surfaces is different for different
phase values. Provided that density fluctuations are negligible, which is
always the case in the strongly degenerate regime \cite{petr2004},  the integrated
interference contrast is completely determined by the shape of the iso-phase
surfaces.

The interference amplitude in a given run of the experiment can be extracted 
from the measured density $n(\vec{x})$
by taking the Fourier transform along $z$ and evaluating it at the wave
vector $Q$, where the magnitude of the local interference amplitude 
\begin{equation}
	A_Q(x,y) = \int\d z\, n(\vec{x})e^{-iQz}
	\label{eq:defAQ}
\end{equation}
takes its maximum. This yields a complex number which contains the random
relative phase between the two clouds. Its average over many runs will
therefore vanish.  Here we are interested in the modulus square of $A_Q$, which
determines the observed visibility of the interference fringes. Experimentally,
the local amplitude $|A_Q(x,y)|^2$ is not a directly accessible quantity since
the absorption imaging automatically integrates over the $y$ direction. In
practice, the averaged interference contrast is obtained by extending the
domain of integration over a volume $\Omega$ that typically covers the entire
sample.  It is then convenient to define an operator
\begin{equation}
	\hat{\alpha} = \int_\Omega\d^3x\int_\Omega\d^3x'\,
	\hat{n}(\vec{x})\hat{n}(\vec{x'})e^{-iQ(z-z')}\ ,
	\label{eq:defalpha}
\end{equation}
whose eigenvalues represent the magnitude of the integrated contrast in an
individual run \cite{polk2006}. In terms of the basic in-trap field operators,
this operator can be expressed by 
\begin{equation}
	\hat{\alpha}=\int_\Omega\d^dr\d^dr'\,
	\hat{\psi}^\dagger_1(\vec{r})\hat{\psi}_1(\vec{r'})
	\hat{\psi}_2(\vec{r})\hat{\psi}^\dagger_2(\vec{r'})\ .
	\label{eq:alphaexplicit}
\end{equation}
Now, in a homogeneous condensate, the one-body density 
matrix $\hat{\psi}^\dagger_i(\vec{r})\hat{\psi}_i(\vec{r'})$ approaches  
a constant on length scales $|\vec{r}-\vec{r'}|$ larger than the healing 
length. Moreover, one has $\hat{\psi}_2(\vec{r})
\hat{\psi}^\dagger_2(\vec{r'})\approx\hat{\psi}^\dagger_2(\vec{r'})
\hat{\psi}_2(\vec{r})$ up to corrections that vanish like $1/\Omega$. 
As a result, the operator 
\begin{multline}
	\hat{\alpha}\simeq\Omega^{-2}
	\left(\int\d^dr\d^dr'\,\hat{\psi}^\dagger_1(\vec{r})
	\hat{\psi}_1(\vec{r'})  \right) 
	\\
	\times\left(\int\d^dr\d^dr'\,\hat{\psi}^\dagger_2(\vec{r})
	\hat{\psi}_2(\vec{r'})  \right) = \hat{N}_0^{(1)}\hat{N}_0^{(2)}\ ,
	\label{eq:alphan0}
\end{multline}
whose eigenvalues determine the measured interference contrast in a given 
run, is equal to the product of the number of condensed atoms
\begin{equation}
	\hat{N}_0^{(i)}=\Omega^{-1}\int\d^dr\d^dr'\,\hat{\psi}^\dagger_i(\vec{r})
\hat{\psi}_i(\vec{r'})
	\label{eq:Nzero}
\end{equation}
within the integration volume $\Omega$ of each initial condensate. 
It is important to emphasize that this argument does not rely on the 
presence of true long range phase coherence. In particular, it is valid
for 2d Bose gases at finite and 1d Bose gases at zero temperature, where
the one particle density matrix approaches a finite value $\tilde{n}_0$ 
on scales much larger than the interparticle spacing. The eventual
algebraic decay to zero only appears at distances beyond a phase 
coherence length $\ell_{\phi}$ that is still much larger.  
For an integration volume that contains a large number of particles $N\gg 1$
and identically prepared samples, therefore, the operator 
$\hat{\alpha}$ is  just the square of the condensed atom number 
in each sample.  The eigenvalues $\alpha$ of $\hat{\alpha}$, which are the 
experimental observables according to the standard rules of quantum mechanics, 
thus may take any value between zero and $N^2$, where $N$ is the number of 
atoms in either of the two samples. The measured integrated density as a function 
of $z$ varies between $2N-2\sqrt{\alpha}$ and $2N+2\sqrt{\alpha}$. The
visibility $\V$ is therefore simply $\sqrt{\alpha}/N$, or
\begin{equation}
	\V^2 = \frac{\alpha}{N^2} = \left( \frac{N_0}{N} \right)^2\ .
	\label{eq:alphavis}
\end{equation}
The measured distribution of the visibility thus directly reflects 
that of the condensate fraction.

Our aim in the following is to calculate the probability distribution for the different positive
eigenvalues $\alpha$ of $\hat{\alpha}$.  For a given many-body
state characterized by a density operator $\hat{\rho}$, the associated 
probability distribution is just 
$p(\alpha)=\langle\alpha|\hat{\rho}|\alpha\rangle$. Mathematically,
it is more convenient to calculate its characteristic function
\begin{equation}
	p(\sigma)\equiv \int_{-\infty}^\infty\d\alpha\,p(\alpha)
	e^{i\sigma\alpha}
	=\Tr\left[ \hat{\rho}e^{i\sigma\hat{\alpha}} \right]
	=\left\langle e^{i\sigma\hat{\alpha}}\right\rangle\ .
	\label{eq:defpsigma}
\end{equation}
To proceed further, we replace  $e^{i\sigma\hat{\alpha}}$ by its normal-ordered
counterpart $\normorder{e^{i\sigma\hat{\alpha}}}$. This approximation neglects 
commutator terms which describe the effect of atomic shot
noise \cite{polk2007} and are of relative order $1/N$ \cite{imam2006}.
Since typical integration volumes for a measurement of the 
interference contrast contain $N\sim 10^3$ atoms at least, 
this approximation is valid up to corrections of less than a percent. 
A convenient representation for the calculation of the characteristic function
\eqref{eq:defpsigma} is obtained by evaluating the trace in terms of 
coherent states. This gives rise to a functional integral
\begin{multline}
	p(\sigma)=(\N_1\N_2)^{-1}
	\int\D(\bar{\psi}_1,\psi_1)\D(\bar{\psi}_2,\psi_2)\\
	\times e^{-S_1[\bar{\psi}_1,\psi_1]-S_2[\bar{\psi}_2,\psi_2]}
	\exp\left[ i\sigma\abs{\int\d^dr\,\bar{\psi}_1(\vec{r})
	\psi_2(\vec{r})}^2 
	\right]
	\label{eq:functint}
\end{multline}
over bosonic c-number fields $\psi(\vec{r},\tau)$, which are periodic in the
interval $\tau=[0,\beta]$ (Here and in the following, we adopt units in which
$\hbar=k_{\text{B}}=1$).  Here $S_{1,2}$ are the respective actions for the
interacting Bose gases $1,2$, while
$\N_i=\int\D(\bar{\psi}_i,\psi_i)\exp(-S_i[\bar{\psi}_i,\psi_i])$ are
normalization factors. Due to our normal-ordering approximation, the fields in
the last exponential do not vary with $\tau$, in contrast to the fields
appearing in the action, but are evaluated at $\tau=0$. For notational
simplicity, $\psi_i(\vec{r},0)\equiv\psi_i(\vec{r})$. By a simple redefinition
of $\sigma\to\sigma/N^2$, Eq.\,\eqref{eq:functint} gives the characteristic
function for the square $\V^2$ of the visibility which is a direct measure of
the interference contrast.

Within the functional integral, it is convenient to switch to the 
density-phase representation
$\psi_i(\vec{r})=\sqrt{n_i(\vec{r})}e^{i\varphi_i(\vec{r})}$ for the c-number fields. 
In this representation, the square of the visibility $\V^2$ reads
\begin{multline}
	\V^2 = \frac{1}{N^2}\int\d^dr\d^dr'\,
	\bar{\psi}_1(\vec{r})\psi_1(\vec{r'})
	\psi_2(\vec{r})\bar{\psi}_2(\vec{r'})
	\\
	= \frac{1}{N^2}\int\d^dr\d^dr'\,
	\sqrt{n_1(\vec{r})n_1(\vec{r'})n_2(\vec{r})n_2(\vec{r'})}
	\\ \times
	e^{i\{[\varphi_2(\vec{r})-\varphi_1(\vec{r})]
	-[\varphi_2(\vec{r'})-\varphi_1(\vec{r'})]\}}\ .
\end{multline}
In particular,  for temperatures low enough that the 
influence of density fluctuations around an average $\bar{n}(\vec{r})$
may be neglected, the visibility 
\begin{equation}
	\V^2 \approx \abs{\frac{1}{N}\int\d^dr\,\bar{n}(\vec{r})
	e^{i[\varphi_2(\vec{r})-\varphi_1(\vec{r})]}}^2
	\label{eq:phasealpha}
\end{equation}
only depends on the 
phase difference $\phi\equiv \varphi_2-\varphi_1$.

\section{Interference contrast at low temperature }
\label{sec:gaussian}

In the following, we will assume that the two gases are identical and describe
each one using the quantum hydrodynamic action
\begin{multline}
	S[\varphi_j]=\int_0^\beta\d\tau\int\d^dr\left\{ 
	\frac{\rhos(\vec{r})}{2m}[\nabla\varphi_j(\vec{r},\tau)]^2
	\right. \\ \left.
	+\frac{1}{2g}[\partial_\tau\varphi_j(\vec{r},\tau)]^2\right\}\ .
	\label{eq:action}
\end{multline}
Here $\beta=1/T$ is the inverse temperature, $m$ is the atomic mass and $g$ is a
coupling constant, which is just the inverse of the compressibility $\kappa$.
Moreover, $\rhos$ is the superfluid
density, which is inhomogeneous in trapped gases. The action
\eqref{eq:action} provides a completely general low-energy description
of superfluid Bose gases. In particular,  it describes 3d gases below the 
critical temperature for Bose-Einstein condensation, 2d gases below the 
Berezinskii-Kosterlitz-Thouless transition \cite{kost1973,bere1971} and 
also 1d gases at zero temperature. 

Since $\V^2$ depends only on the phase difference, it is advantageous to
switch to a new set of variables:
\begin{equation}
	\Phi = \frac{\varphi_2 + \varphi_1}{2}
	\quad ;\quad
	\phi = \varphi_2-\varphi_1\ .
	\label{eq:pmvars}
\end{equation}
In terms of these variables, the total action can be rearranged as
\begin{equation}
	S = S[\varphi_1]+S[\varphi_2]
	= 2S[\Phi]+\frac{1}{2}S[\phi]\ ,
	\label{eq:pmaction}
\end{equation}
which implies that the contribution depending on the total average phase 
$\Phi$ cancels out in equation \eqref{eq:functint}.
The characteristic function \eqref{eq:functint} can then be written as
\begin{equation}
	p(\sigma)=\frac{1}{\N}\int\D\phi\, e^{-S[\phi]/2}	
	\exp\left[ i\sigma\abs{\int\frac{\d^dr}{N}\,\bar{n}(\vec{r})
	e^{i\phi(\vec{r})}}^2 \right]\ ,
	\label{eq:finalpomega}
\end{equation}
with $\N=\int\D\phi\,\exp(-S[\phi]/2)$. Note that
a constant contribution in $\phi$ has no effect on the result since it 
cancels out when taking the modulus square. In the following, we will thus
restrict our analysis to functions without constant component, \ie,
$\int\d^dr\,\phi(\vec{r})=0$.

For explicit calculations, we follow the technique used by Imambekov \etal
\cite{imam2006,imam2008} and parametrize the functional integral
\eqref{eq:finalpomega} by expanding $\phi$ in terms of the solutions of the imaginary
time Euler-Lagrange equation associated with the action \eqref{eq:action},
\begin{equation}
	\partial_\tau^2\phi(\vec{r},\tau) + \frac{g}{m}\nabla\cdot\left[ 
	\rhos(\vec{r})\nabla\phi(\vec{r},\tau)\right] = 0\ ,
	\label{eq:eulerlagrange}
\end{equation}
supplemented with appropriate spatial boundary conditions. The bosonic nature
of the field $\phi(\vec{r},\tau)$ requires that the solutions satisfy
$\phi(\vec{r},0)=\phi(\vec{r},\beta)$. A separation ansatz readily gives a
family of solutions $\psi_\lambda(\vec{r})e^{\pm\omega_\lambda\tau}$, where
$\lambda$ is a formal index labelling the eigenmodes. To satisfy the boundary
condition on $\tau$, we use the expansion
\begin{equation}
	\phi(\vec{r},\tau)=\sum_{\lambda\neq 0}
	s_\lambda\psi_\lambda(\vec{r})\left( 
	\frac{e^{\omega_\lambda\tau}}{e^{\beta\omega_\lambda}+1}
	+\frac{e^{\beta\omega_\lambda}e^{-\omega_\lambda\tau}
	}{e^{\beta\omega_\lambda}+1}\right)\ .
	\label{eq:phiexpansion}
\end{equation}
Factors have been chosen so that the parentheses evaluate to unity at $\tau=0$
and $\tau=\beta$.

Unless excluded by the boundary conditions, the Euler-Lagrange equation always
permits a solution $\psi_0(\vec{r})$ which is constant and non-zero in space.
Since the modes $\psi_\lambda$ form a complete orthonormal system,
$\int\d^dr\,\psi_\lambda(\vec{r})\psi_{\lambda'}(\vec{r})
=\delta_{\lambda,\lambda'}$, this implies that all further modes have a
vanishing spatial average. The condition $\int\d^dr\,\phi=0$ therefore
translates to the omission of the constant mode $\lambda=0$. 

Substituting the expansion \eqref{eq:phiexpansion} into the action yields
a diagonal quadratic form
\begin{equation}
	\frac{S}{2}=\sum_{\lambda\neq 0}\frac{s_\lambda^2}{2g}
	\omega_\lambda\tanh\left( \frac{\beta\omega_\lambda}{2}\right)\ .
	\label{eq:expandaction}
\end{equation}
Introducing the dimensionless variables 
\begin{equation}
	t_\lambda\equiv 
	s_\lambda\sqrt{\frac{\omega_\lambda}{g}\tanh\left(
	\frac{\beta\omega_\lambda}{2}\right)}\ , 
	\label{eq:deft}
\end{equation}
the characteristic function can then be rewritten as
\begin{equation}
	p(\sigma)=\prod_{\lambda\neq 0}\int\frac{\d t_{\lambda}\, 
	e^{-\frac{t_{\lambda}^2}{2}}}{\sqrt{2\pi}}
	\exp\left( i\sigma\left|\int\frac{\d^dr}{N}\,\bar{n}(\vec{r})
	e^{ih_{\{t_{\lambda}\}}(\vec{r})}\right|^2 \right)
	\label{eq:pomegaexpand}
\end{equation}
where
\begin{equation}
	h_{\{t_{\lambda}\}}(\vec{r})=\phi(\vec{r},0)=\sum_{\lambda\neq 0}
	\sqrt{\frac{g}{\omega_{\lambda}}\coth\left( 
	\frac{\beta\omega_\lambda}{2}\right)}t_{\lambda}
	\psi_{\lambda}(\vec{r})
	\label{eq:height}
\end{equation}
is the parametrized iso-phase surface.  This expression has already been
derived  by Imambekov \etal \cite{imam2008}, starting from the
relationship between the moments of $\V^2$ and higher-order correlation
functions. The fluctuating surface $h$ emerged as an abstraction in their
paper. Here, we see that it is just the shape of the iso-phase
surfaces.

Eqs.\,\eqref{eq:pomegaexpand} (or rather, the expression for $\V^2$ appearing
therein) and \eqref{eq:height} are well suited for numerical use: to obtain
a given realization of $\V^2$, one generates a large number 
of Gaussian deviates for the
amplitudes $t_\lambda$, constructs the corresponding surface $h$ and numerically
calculates the integral to obtain $\V^2$. Repeating this sequence with
different sets of amplitudes yields histograms for the possible
values of $\V^2$, whose shape approaches the actual probability
distribution as the number of iterations grows large. 

Analytical results can be obtained in the limit of high contrasts,
$1-\V^2\ll 1$, where the iso-phase surfaces are
smooth and one may expand the exponential inside the definition of $\V^2$.
As is clear from equation \eqref{eq:height}, the expansion parameter is
given by
\begin{equation}
	\epsilon\equiv\max_{\lambda,\vec{r}}
	\sqrt{\frac{g}{\omega_\lambda}\coth\left( 
	\frac{\beta\omega_\lambda}{2} \right)}
	\psi_{\lambda}(\vec{r})\ .
	\label{eq:defeps}
\end{equation}
In all cases that we will encounter in the following, this maximum is
found for $\omega_{\text{min}}\equiv\min_{\lambda\neq 0}\omega_\lambda$.

In the following, we will focus on the regime where the reduction of the
visibility is dominated by thermal fluctuations so that quantum fluctuations
(caused by interactions) may be neglected. Considering, in particular, a 2d
Bose gas whose characteristic size $l_z$ in the direction of transverse
confinement obeys $l_z\gg a_s$, the effect of zero point fluctuations of the
phase on the reduction of the visibility can be determined by a Bogoliubov
calculation, which gives \cite{petr2004}
\begin{equation}
	\langle\V^2\rangle(T=0)=\left(\frac{n_0(0)}{n}\right)^2
	=1-\frac{\tilde{g}_2}{2\pi}+
	\cdots \ .
	\label{eq:dep2dt0}
\end{equation}
Here, $\tilde{g}_2=mg_2=\sqrt{8\pi}a_s/l_z$ is the dimensionless 2d interaction 
constant, which has typical values $\tilde{g}=0.1$ \cite{hadz2006,hadzunpu},
which implies practically unit visibility at zero temperature. By
contrast, at finite temperature, the integrated visibility 
\begin{equation}
	\langle\V^2\rangle(T)-\langle\V^2\rangle(0)=
	\left[ 1-2\eta(T)\log\left( \frac{L}{\xi} \right) \right]
	\label{eq:exp2d}
\end{equation}
decreases logarithmically with the size of the system, which reflects
the absence of long range order at finite temperature in the
thermodynamic limit \cite{hohe1967,merm1966}.  Since the size
dependence of the thermal depletion is only logarithmic, finite condensate
fractions may be found at low enough temperature for realistic system sizes.

In actual experiments, the temperature is usually large compared to
the typical frequencies $\omega_\lambda$ which are of the order of the chemical
potential $\mu$.  Then
$\coth(\beta\omega_\lambda/2)$ can be replaced by $2T/\omega_\lambda$ and hence
$\epsilon^2=2gT\max_{\lambda,\vec{r}}\psi_\lambda(\vec{r})^2/\omega_\lambda^2$.
Within this approximation the effect of thermal phase fluctuations scales
linearly with temperature. For weakly interacting 2d Bose gases, the necessary
condition $T\gg\mu$ is well obeyed even in the deeply degenerate regime because
$\mu/T=\tilde{g}_2/2\pi$ at $n\lambda_T^2=\mathcal{O}(1)$ \cite{bloc2008}.
Note that for $\tilde{g}_2\ll 1$ the conditions $T\gg\mu$ and $\epsilon^2\ll 1$
are simultaneously satisfied for the typical phase space densities 
$n\lambda_T^2$ that are reached in 2d Bose gases \cite{hadz2009}.

Quite generally, whether or not $\epsilon$ is small, \ie, whether a physical
regime which permits such an expansion exists, depends crucially on temperature
and the spatial dimension $d$.  Because of normalization, the eigenfunctions
$\psi_\lambda$ scale with the characteristic size $L=\Omega^{1/d}$ of the
system as $L^{-d/2}$, while the eigenfrequencies $\omega_\lambda$ will scale as
$L^{-1}$, independent of the dimensionality. It follows that $\epsilon$ scales
as $L^{1-d/2}$.  For concreteness, in a homogeneous system with periodic
boundary conditions,
\begin{equation}
	\epsilon^2=\frac{L^{2-d}}{\pi^2}\frac{mT}{\rhos}\, .
	\label{eq:epsclass}
\end{equation}  
For the case of 2d Bose gases, which is the main focus of our work,
$\epsilon^2=2\eta(T)/\pi$ is independent of  system size and is
determined by the exponent $\eta(T)=(\rhos\lambda_T^2)^{-1}$ which gives the
decay of the one-body density matrix. 

For $\epsilon\ll 1$, the exponent in equation \eqref{eq:pomegaexpand} can 
be expanded in the form
\begin{equation}
\exp\Bigg(i\sum_{\lambda\neq 0}\epsilon t_\lambda\Bigg)
\simeq 1+i\sum_{\lambda\neq 0}\epsilon t_\lambda
-\frac{1}{2}\sum_{\lambda,\lambda'\neq 0}\epsilon^2 t_\lambda t_{\lambda'}
\label{eq:spinwave}
\end{equation}
since the variables $t_\lambda$ are of order one due to the Gaussian weight
factors $e^{-t_\lambda^2/2}$. Within this approximation, an exact calculation
of the distribution functions is possible. The inclusion of the terms quadratic
in $\epsilon$ leads to non-trivial distributions instead of the Delta functions
that result in leading order in $\epsilon$ \cite{imam2006}.  In the general
case of an inhomogeneous system with a spatially varying superfluid density
$n_s(\vec{r})$, thermal phase fluctuations lead to a reduction
of the visibility from unity (or---more precisely---from its value at zero
temperature) of the form
\begin{equation}
	1-\V^2=\frac{2gT}{\Omega}\left[ 
	\sum_{\lambda,\lambda'\neq 0}\frac{t_\lambda t_{\lambda'}}
	{\omega_\lambda\omega_{\lambda'}}I_{\lambda,\lambda'}
	-\left( \sum_{\lambda\neq 0}\frac{t_\lambda}{\omega_\lambda} 
	J_\lambda
	\right)^2
	\right]\ ,
	\label{eq:firstorder}
\end{equation}
where $\Omega$ is the integration volume  and
\begin{equation}
	\begin{split}
	I_{\lambda,\lambda'}&=\frac{\Omega}{N}\int\d^dr\,\bar{n}(\vec{r})
	\psi_\lambda(\vec{r})\psi_{\lambda'}(\vec{r})\ ,
	\\
	J_\lambda&= \frac{\sqrt{\Omega}}{N}\int\d^dr\,\bar{n}(\vec{r})
	\psi_\lambda(\vec{r})
\end{split}
	\label{eq:integrals}
\end{equation}
are dimensionless numbers.
In a homogeneous system, $I_{\lambda,\lambda'}=\delta_{\lambda,\lambda'}$,
and $J_\lambda=0$ for any $\lambda$. In inhomogeneous systems, there may
be finite ``off-diagonal'' values for $I_{\lambda,\lambda'}$ and finite values 
for $J_\lambda$.

For a discussion of some general features of the statistics of the interference contrast
like the dependence on dimensionality and the related issue of self-averaging, 
we focus on homogeneous systems (we will discuss the 
experimentally relevant  trapped 2d system in section \ref{sec:ho}).
It is then convenient to define
\begin{equation}
	u \equiv\frac{2}{\epsilon^2}(1-\V^2)
	=\sum_{\lambda\neq 0}\frac{t_\lambda^2}{\tilde{\omega}_\lambda^2}
	\ ,
	\label{eq:defu}
\end{equation}
where $\epsilon^2$ has been defined in equation \eqref{eq:epsclass} and
$\tilde{\omega}_\lambda \equiv \omega_\lambda/\omega_{\text{min}}$. Note
that the scaling factor $2/\epsilon^2$ between $1-\V^2$ and $u$ is 
large compared to one. While the visibility
takes values on the interval $[0,1]$, the auxiliary variable $u$ has values on
the interval $[0,\infty]$ due to our expansion of $e^{i\phi(\vec{r})}$.  The
characteristic function
\begin{equation}
	q(\sigma)=\langle e^{i\sigma u}\rangle =
	\int\prod_{\lambda\neq 0}\frac{\d t_{\lambda}\, 
	e^{-t_{\lambda}^2/2}}{\sqrt{2\pi}}
	\exp\left(i\sigma\sum_{\lambda\neq 0}
	\frac{t_{\lambda}^2}{\tilde{\omega}_\lambda^2}  \right)
	\label{eq:qomegau}
\end{equation}
of the probability distribution $q(u)$ for the rescaled deviation $u$ of the 
visibility from unity is now readily evaluated to be
\begin{equation}
	q(\sigma)=\prod_{\lambda\neq 0}\frac{1}
	{\sqrt{1-2i\sigma/\tilde{\omega}_\lambda^2}}\ .
	\label{eq:charfunction}
\end{equation}
This evaluation ceases to be straightforward when $I_{\lambda,\lambda'}$ is
not diagonal, since it amounts to the calculation of the determinant of an
infinite matrix with a non-trivial entry structure. 

The logarithm 
\begin{equation}
	\log q(\sigma)=\frac{1}{2}\sum_{s=1}^\infty
	\frac{(2i\sigma)^s}{s}\zeta_{\{\lambda\}}(s)
	=\sum_{s=1}^\infty\frac{(i\sigma)^s}{s!}\langle u^s\rangle_\text{c}
	\label{eq:defcumgen}
\end{equation}
of the characteristic function $q(\sigma)$, which is the generating function of the 
cumulants of $u$, can be expressed in terms of the spectral zeta function 
$\zeta_{\{\lambda\}}(s)\equiv\sum_{\lambda\neq 0}(\tilde{\omega}_\lambda^2)^{-s}$
of the eigenfrequencies of the quantum hydrodynamic action \eqref{eq:action}.
In particular, it determines all cumulants of the random variable $u$ via
\begin{equation}
	\langle u^s\rangle_\text{c}= 2^{s-1}(s-1)!\,\zeta_{\{\lambda\}}(s)\ .
	\label{eq:cumulants}
\end{equation}
Note that this calculation does not depend on the explicit form of the
eigenvalues and eigenfunctions: the geometry of the system is completely
contained in the factor  between $u$ and $1-\V^2$ and the spectral zeta
function (for systems with diagonal $I_{\lambda,\lambda'}$).

The precise form of the spectral zeta function obviously depends on the
geometry of the system and the spectrum that follows from it, but the following
properties are valid for any homogeneous system: 
$\zeta_{\{\lambda\}}(s)$ is a monotonically decreasing function of its argument
and has a lower bound (which is reached in the limit $s\rightarrow\infty$)
equal to the degeneracy of $\omega_{\text{min}}$ (note that this essential
property cannot be reproduced when one replaces the sum in 
$\zeta_{\{\lambda\}}(s)$ by an integral). It follows that for
increasing $s$, the number of frequencies that make a non-negligible 
contribution to the value of $\zeta_{\left\{ \lambda \right\}}(s)$ decreases
so that higher-order cumulants will essentially depend only on a small number
of low frequencies. However, we will find that it diverges for $s=1$ in 
$d\geq 2$ and must be rendered finite by the introduction of a UV cutoff.

Substituting back from equation \eqref{eq:defu}, we obtain the cumulants of 
$\V^2-1$ (which are identical to the cumulants of $\V^2$ except for the 
expectation):
\begin{equation}
	\langle(\V^2-1)^s\rangle_\text{c}= 
	\frac{(s-1)!}{2}\left(-\epsilon^2\right)^s
	\zeta_{\{\lambda\}}(s)\ .
	\label{eq:cumalpha}
\end{equation}
As we will see, the spectral zeta function remains finite for all $s\geq 2$
in all relevant cases, thus determining the finite size scaling behavior of all 
higher cumulants: in $d$ dimensions, the $s$th cumulant scales as $L^{s(2-d)}$. 

Specifically, we consider a homogeneous system in $d$ dimensions in
a hypercubic volume $\Omega=L^d$ with periodic boundary conditions. Then, 
$\omega_{\vec{k}}=c|\vec{k}|$, with $\vec{k}=(2\pi/L)(l_1,\dots,l_d)$,
where $l_i\in\mathbb{Z}$, and
the speed of sound $c=\sqrt{g\rhos/m}$. The resulting spectral zeta function then reads
\begin{equation}	
	\zeta_{\{\vec{k}\}}(s)={\sum_{l_1,\dots,l_d}}'
	\frac{1}{(l_1^2+\cdots+l_d^2)^s}
	=\sum_{n=1}^\infty A_d(n)\frac{1}{n^s}	\ ,
	\label{eq:pbczeta}
\end{equation}
where the prime on the sum indicates that the point $l_1=\cdots=l_d=0$ is
omitted and $A_d(n)$ is the number of possibilities to represent the integer
$n$ as a sum of $d$ squares (including squares of negative numbers). The
representation on the right-hand side is a special case of a Dirichlet series
which arises in connections between number-theory and modular forms
\cite{frei2008}.

The expectation is then given by equations \eqref{eq:cumalpha} and
\eqref{eq:pbczeta} with $s=1$. However, simple power counting reveals that the
spectral zeta function is ultraviolet (UV) divergent at $s=1$ in two and three
dimensions. Since the quantum hydrodynamic action \eqref{eq:action} is an
effective low-energy description, however, this divergence is an artifact. It
can be avoided by introducing a cutoff at a maximum momentum
$\Lambda=2\pi/\xi$, where $\xi$ is the healing length.  The necessity of an
explicit UV cutoff has the important consequence that the expectation does not
follow the scaling with system size announced in equation \eqref{eq:cumalpha}:
in 2d, the cutoff introduces a logarithmic dependence on system size that would
otherwise be absent so that the expectation takes on a non-universal character.
During the remainder of this article, we will focus on the higher cumulants
(from the variance on) which are universal.

Since the variance is independent of system size (as are all higher cumulants)
while the expectation vanishes logarithmically as $L\rightarrow\infty$,
fluctuations are not self-averaging in 2d and the regime $1-\V^2\ll 1$ can only
be reached in finite-size systems (even if they may in fact be quite large due
to the weak logarithmic size dependence of the expectation). This is in
contrast to the 3d situation where as a consequence of true long range order,
the expectation is finite in the thermodynamic limit while all higher cumulants
decrease with increasing system size. In this case, the visibility is a
self-averaging observable, \ie, for large integration volumes $\Omega$, the
value obtained in a single run is equal to an average over many runs.
In section~\ref{sec:analyt}, we will
focus on system sizes where the condensate depletion remains small and the
visibility is close to unity, \ie, $L\ll \ell_\phi = \xi e^{1/2\eta(T)}$ and
the system is a true condensate rather than a quasi-condensate \cite{petr2004}.
The opposite case will be discussed in section \ref{sec:critpoint}.

The fact that all cumulants from the variance on are finite and obey the
scaling given in equation \eqref{eq:cumalpha} implies that the probability
distribution $p(\V^2)$ is universal and non-Gaussian. The following section is
devoted to the explicit analytic calculation of this distribution in different
geometries.  

\section{Analytical results in 2d}
\label{sec:analyt}

For a given geometry and boundary conditions, one can always evaluate the 
spectral zeta function numerically. Equation \eqref{eq:cumalpha} 
then permits the calculation of an arbitrary
number of cumulants. However, this is
not sufficient to obtain an analytical expression for the underlying
probability distribution $p(\V^2)$ which requires a closed-form expression
for $\zeta_{\{\lambda\}}(s)$.  This may be obtained in the 2d case for
simple geometries, which we will discuss in this section. 

In order to clarify the applicability of our results to experiment, it is
necessary to define what we mean by ``2d'' in practice. We consider the two
gases to be strongly confined along the $z$ direction with a trapping potential
$\omega_z$ which is sufficiently strong to ensure $\omega_z\gg T$ and
$\omega_z\gg\mu$ so that the gases reside in the harmonic oscillator ground
state along the $z$ direction. However, the spatial extension
$l_z=\sqrt{1/m\omega_z}$ shall still be large compared to the scattering length
$a_s$. This regime, which is sometimes referred to as ``quasi-2d''
\cite{petr2001}, accurately describes the situation in typical experiments on
cold atoms where the tight confinement is realized using an optical dipole
potential \cite{hadz2006,krug2007,clad2009,gill2009,rath2010,hung2010}.  It is
particularly simple in that the interaction may be described by a simple
dimensionless constant $\tilde{g}=mg=\sqrt{8\pi}a_s/l_z$ (for not too strong
interactions. Here and in the following, we drop the subscript ``2'' for ease
of notation) whereas the interaction constant in a two-dimensional system
with $l_z\lesssim a_s$ depends on the chemical potential and thus effectively
on the spatial density \cite{petr2001}.

The first case we discuss is that of a rectangle with periodic boundary
conditions. The latter are certainly artificial, but the results are important
from a conceptual point of view because the calculation can be carried out in
closed form.  We will discuss two limiting cases of this particular case
(isotropic and strongly anisotropic) before moving on to the experimentally
relevant case of a harmonically trapped sample.

\subsection{Homogeneous square sample}
\label{ss:homsq}

The conceptually and mathematically most elementary case is the one where
the gas is homogeneous and confined to a rectangular area of extension $L\times
aL$, $0<a\leq 1$, with periodic boundary conditions. First, we will focus
on the particular case $a=1$, \ie, a square sample, but 
it is convenient to introduce the notation directly in its slightly more
general form for arbitrary $a$.  In this case, the eigenfunctions take the
simple form
\begin{equation}
	\psi_{\vec{k}}^{(c)}=\sqrt{\frac{2}{aL^2}}\cos(\vec{k}\cdot\vec{r})
	\quad ;\quad
	\psi_{\vec{k}}^{(s)}=\sqrt{\frac{2}{aL^2}}\sin(\vec{k}\cdot\vec{r})
	\label{eq:psipbc}
\end{equation}
with $\vec{k}=(2\pi/L)\times(n_1/a,n_2)$, and $n_{1,2}\in\mathbb{Z}$. 

We may now give an explicit expression for the expansion parameter $\epsilon$
defined in equation \eqref{eq:defeps} and the relationship between $\V^2$
and the auxiliary variable $u$. For this particular set of
eigenfunctions and eigenvalues,
\begin{equation}
	\epsilon^2=\frac{1}{a\pi^2}\frac{mT}{\rhos}=\frac{2}{\pi a}\eta(T)
	\quad ;\quad
	u=\frac{2}{\epsilon^2}(1-\V^2)\ .
\label{eq:homepsilon}
\end{equation}
Quite remarkably, $\epsilon$ and hence the entire probability distribution
$p(\V^2)$ have no explicit dependence on the interaction constant $g$, which 
is hidden in the non-trivial relation between
the superfluid density that enters $\eta(T)$ and the bare 2d density $n$.
For weakly interacting Bose gases, this relation has been worked out in 
\cite{prok2002}.

As we already stated, the calculation of a closed-form expression for the
probability distribution $p(\V^2)$ requires a closed-form expression
for \zetanb{a}. It turns out that in two dimensions and in the absence of a
high-$k$ cutoff \zetanb{1} is a special case of a lattice sum first calculated 
by Lorenz \cite{lore1871} and Hardy \cite{hard1919} who showed that
\begin{equation}
	\zetanb{1}(s)={\sum_{l_1,l_2}}'\frac{1}{(l_1^2+l_2^2)^s}
	=4\zeta(s)\beta(s)\ ,
\label{eq:zetanb}
\end{equation}
where the prime on the sum signifies that the value $l_1=l_2=0$ must be
omitted (for a derivation of equation \eqref{eq:zetanb}, see \cite{zuck1974}).  
Here, $\zeta(s)=\sum_{k=1}^\infty k^{-s}$ is the Riemann zeta function and
$\beta(s)=\sum_{l=0}^\infty(-1)^l/(2l+1)^s$ is the Dirichlet beta function.
As already stated, \eqref{eq:zetanb} is undefined for
$s=1$ since the Riemann zeta function diverges, \ie, the expectation of $u$
cannot be calculated without a UV cutoff. For $s\geq 2$, \ie, for all
cumulants except the expectation, \eqref{eq:zetanb} is finite. The introduction
of a high-$k$ cutoff will only cause small deviations since the sums are
infrared-dominated and we may write 
\begin{equation} \log
	q(\sigma)=i\sigma\langle u\rangle +
	2\sum_{s=2}^\infty\frac{(2i\sigma)^s}{s}\zeta(s)\beta(s) \ ,
	\label{eq:logpiso} 
\end{equation} 
where $\langle u\rangle$ is evaluated using a finite cutoff. The different
UV-behaviour of the expectation and the higher-order cumulants has the
consequence that the probability distribution for $u$ is universal except for a
cutoff-dependent shift. Thus, the variable $u-\langle u\rangle$ has values on
the entire real axis. As long as the resulting probability distribution takes
on negligible values for values of $u$ corresponding to visibilities outside
the interval $[0,1]$, this does not create any inconsistency.

As pointed out by Bramwell in the context of the order parameter distribution
for the 2d XY model in the low temperature limit \cite{bram2009}, 
\eqref{eq:logpiso} is closely related to the Gumbel distribution 
\begin{equation}
p_{\text{G}}(x)=\exp[-(x+\gamma)-e^{-(x+\gamma)}]
\label{eq:pxgumbel}
\end{equation}
(with the Euler constant $\gamma = 0.577\dots$) which determines the 
statistics of the interference contrast for weakly interacting 1d Bose 
gases at zero temperature \cite{grit2006,imam2006} 
(note that \eqref{eq:pxgumbel} is a normalized distribution with zero average 
and variance $\pi^2/6$). Its cumulant generating function reads 
\begin{equation}
\log p_{\text{G}}(\sigma)=\sum_{s=2}^\infty\frac{(i\sigma)^s}{s}\zeta(s)\ .
\label{eq:logpwgumbel}
\end{equation}
In fact, there are two non-trivial differences:
(i) the presence of the
Dirichlet beta function in \eqref{eq:logpiso}. This function rapidly
converges to one so that it may be replaced by unity to
a good approximation (see table \ref{tab:beta}). Since the expectation is
non-universal anyway, this essentially amounts to increasing the variance
by 9\%, or equivalently, the width of the distribution by 4\% (this may be seen
as a convolution with a normalized Gaussian of appropriate 
width). \begin{table}
	\centering
	\begin{tabular}{c|c|c|c|c|c}
		$s$ & $2$ & $3$ & $4$ & $5$ \\
		\hline\hline
		$\beta(s)$ &
$0.915966$ & $0.968946$ & $0.988945$ & $0.996158$
	\end{tabular}
	\caption{The first four relevant values of the Dirichlet beta function}
	\label{tab:beta}
\end{table}
(ii) the global factor of two which implies (accepting $\beta(s)\simeq 1$) that
$q(u)$ is the convolution of two identical Gumbel distributions, and
$p(\V^2)$ its scaled mirror image.  Thus, there is a striking similarity
between the 2d case at finite temperature discussed here and the 1d case at
vanishing temperature. However, owing to the different corresponding
eigenspectra, passing from the latter to the former case does not amount to the
simple replacement $K\mapsto 1/2\eta$ as suggested in \cite{imam2008}.

\begin{figure}[htbp]
	\centering
	\includegraphics[width=\columnwidth]{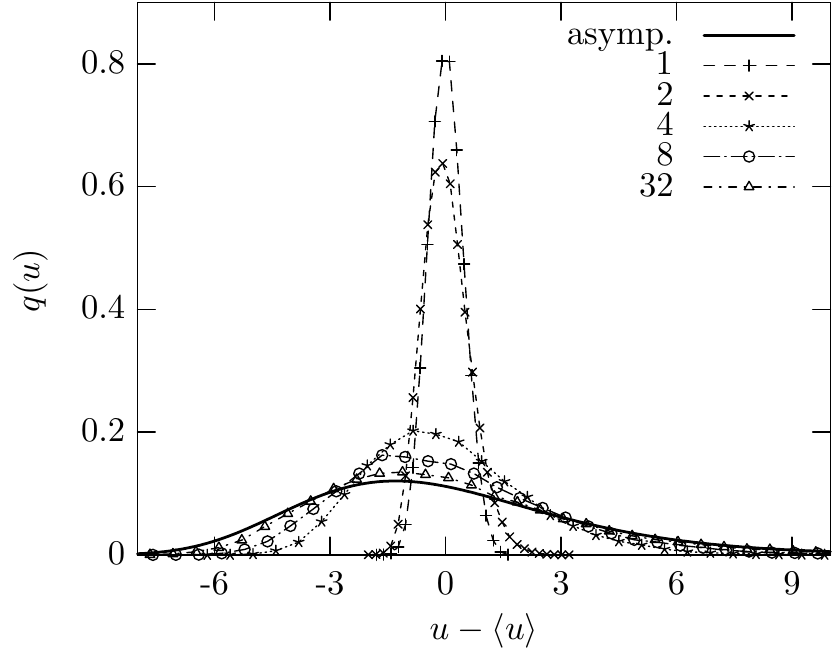}
	\caption{Numerically calculated distributions $q(u-\langle u\rangle)$
	for $T_c/T=1,2,4,8,32$ (symbols, cf.\ legend) and a convolution of two 
	Gumbel 	distributions (continuous line). The lines are guides for
	the eye. Here and in the following plots, the statistical error is of
	the order of the symbol size.}
	\label{fig:u}
\end{figure}
The evolution of the distribution $q(u)$ with decreasing temperature, obtained
numerically, is shown in Fig.\,\ref{fig:u}. For simplicity, we have subtracted
the non-universal expectation so that all distributions are centered around
zero.  The numerical data was obtained by generating 100\,000 random surfaces
per curve using a total of 1000~modes on a 16\,000~point grid and then
calculating $\V^2$ using equations \eqref{eq:pomegaexpand} and
\eqref{eq:height} without any approximation beyond the replacement of
$\coth(\beta\omega_\lambda/2)$ by $2T/\omega_\lambda$. In particular, there is
no expansion of $\exp(ih)$ to second order in $h$. For
sufficiently low temperatures, the distribution approaches the universal low
temperature distribution, whose characteristic function is given in equation
\eqref{eq:logpiso}. The numerically calculated distribution for
$\rhos/mT=2/\pi$ should of course
not be taken seriously: while the superfluid density remains finite at the
critical point, the range of momenta where the quantum hydrodynamic action
\eqref{eq:action} is valid approaches zero. A proper result for the
distribution of the interference contrast near $T_c$ requires calculating the
full counting statistics of the condensate number near the BKT-transition, as
will be discussed in section \ref{sec:critpoint}.  
\begin{figure}[htbp]
	\centering
		\includegraphics[width=\linewidth]{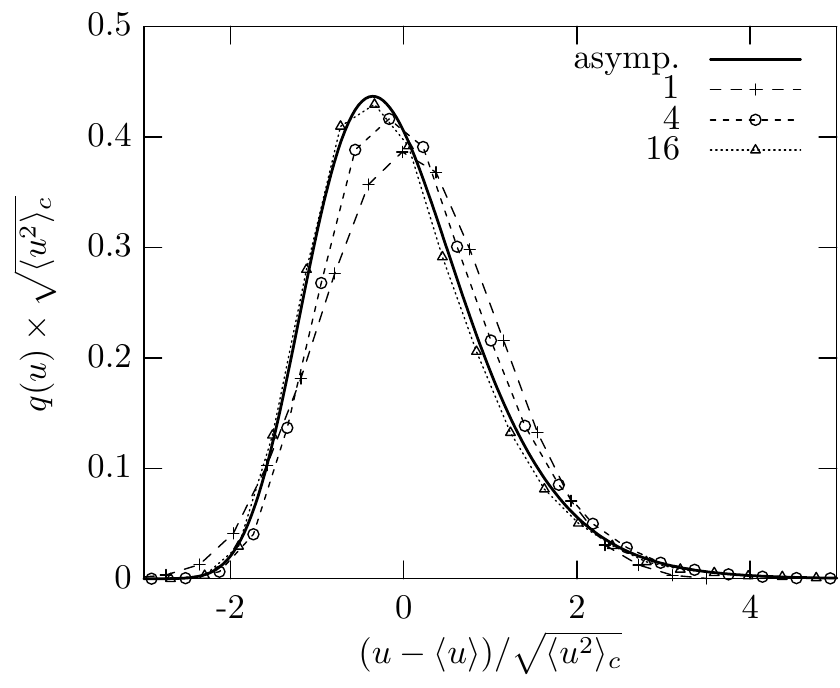}
	\caption{The same data as in the previous figure (except for the choice
	of shown temperatures), but with
	normalised variance. In this representation, the convergence towards
	the asymptotic shape is considerably faster.}
	\label{fig:uscaled}
\end{figure}

As one can see in Fig.\,\ref{fig:uscaled}, the actual shape of the 
distribution that is obtained from scaling $q(u)$ by its standard deviation 
$\sqrt{\langle u^2\rangle_c}$ is already quite close to the asymptotic result 
for $\rhos/mT=2/\pi$ and converges rapidly with decreasing temperature.

\subsection{Strongly anisotropic rectangle}
\label{ss:aniso}

While there is no general closed form expression for \zetanb{a} for arbitrary
values of $a$, it is nonetheless possible to obtain an analytical expression
for $p(\V^2)$ in the limiting case $a\ll 1$, \ie, for a very anisotropic
sample. The reason lies in the mathematical structure of the cumulants: except
for the non-universal expectation, all cumulants are given by 
sums which are dominated by the
lowest-lying eigenvalues. Already for moderate values of the aspect ratio
$a$ (around $1/5$. Preliminary experimental data has been taken for aspect
ratios even lower, $\lesssim 0.1$ \cite{hadzunpu}), the contribution of the 
modes along the shorter direction of the samples becomes negligible and one 
obtains
\begin{equation}
	\zetanb{a\ll 1}(s)\simeq\sum_{l\neq 0}\frac{1}{l^{2s}}
	=2\zeta(2s)\ .
	\label{eq:zetaaniso}
\end{equation}

It is important to keep in mind that in order to remain in the two-dimensional
regime, the aspect ratio must not become too small, \ie, equation 
\eqref{eq:zetaaniso} holds under the condition that $a\ll 1$, but at the same
time $\hbar^2/m(aL)^2 < \mu,T$. If the second inequality is violated, the system
becomes effectively one-dimensional. At the same time, $\epsilon$ as given in
equation \eqref{eq:defeps} ceases to be a small parameter so that one can no
longer justify the approximate treatment of $e^{i\phi(\vec{r})}$. However, we
emphasize that equation \eqref{eq:zetaaniso} is well satisfied (for $s\geq 2$) 
already for moderately small $a$ so that the strongly anisotropic 2d regime
is well-defined.

Substituting \eqref{eq:zetaaniso} into equation \eqref{eq:defcumgen} yields
\begin{multline}
	\log q(\sigma)=i\sigma \langle u\rangle + \sum_{s=2}^\infty
	\frac{(2i\sigma)^s}{s}\zeta(2s) \\
	= i\sigma\left(\langle u\rangle -\frac{\pi^2}{3}\right)+
	\log\left[ \Gamma\left(1-\sqrt{2i\sigma}\right)
	\Gamma\left(1+\sqrt{2i\sigma}\right) \right]
	\ ,
	\label{eq:logpwaniso}
\end{multline}
or (defining $u_{\text{min}}\equiv \langle u\rangle -\pi^2/3$)
\begin{equation}
	q(\sigma) = \pi\sqrt{2i\sigma}\csc\left(\pi\sqrt{2i\sigma}\right)
	e^{i\sigma u_{\text{min}}}\ .
	\label{eq:pwaniso}
\end{equation}
This function is meromorphic in the entire complex plane since the branch
cuts of the square roots before and inside the cosecant cancel each other.
This permits to explicitly calculate its inverse Fourier transform 
$q(u)=(2\pi)^{-1}\int_{-\infty}^\infty\d\sigma\, q(\sigma)e^{-iu\sigma}$ using
the residue theorem. 
In the upper half plane, $q(\sigma)$ has no poles and falls off as
$\exp[-\sqrt{2\/\Im(\sigma)}]$ for large arguments. Thus, for 
$u- u_{\text{min}}\leq 0$, one may close the integration contour with a 
half-circle over the upper half plane and the integral vanishes. For
$u-u_{\text{min}}>0$, one must close the contour in the lower half plane where
$q(\sigma)$ has an infinite number of poles $\sigma_n=-in^2/2$. There,
$e^{-i\sigma(u-u_{\text{min}})}q(\sigma)$ has the residues
$R_n=(-1)^{n-1}in^2e^{-n^2(u-u_{\text{min}})/2}$. Thus we obtain
\begin{equation}
	q(u)=\begin{cases}
		0 & u\leq u_{\text{min}}\\
		\sum_{n=1}^\infty(-1)^{n-1}n^2e^{-\frac{n^2}{2}(u-u_{\text{min}})} & u > u_{\text{min}} 
	\end{cases}\ .
	\label{eq:puaniso}
\end{equation}
This can be written in a more compact form in terms of its cumulative 
distribution function:
\begin{equation}
	q(u) = \theta(u-u_{\text{min}}^+)\frac{\d}{\d u}
	\vartheta_4\left(e^{(u-u_{\text{min}})/2}\right)\ ,
	\label{eq:puanisoshort}
\end{equation}
where $\theta$ is the Heaviside function, and 
$\vartheta_4(z)=1+2\sum_{n=1}^\infty (-1)^n z^{n^2}$ is a Jacobi theta 
function. As a shorthand, we will refer to the distribution described by
equation \eqref{eq:puaniso} as ``Jacobi distribution''.

\begin{figure}[htbp]
	\centering
	\includegraphics[width=\linewidth]{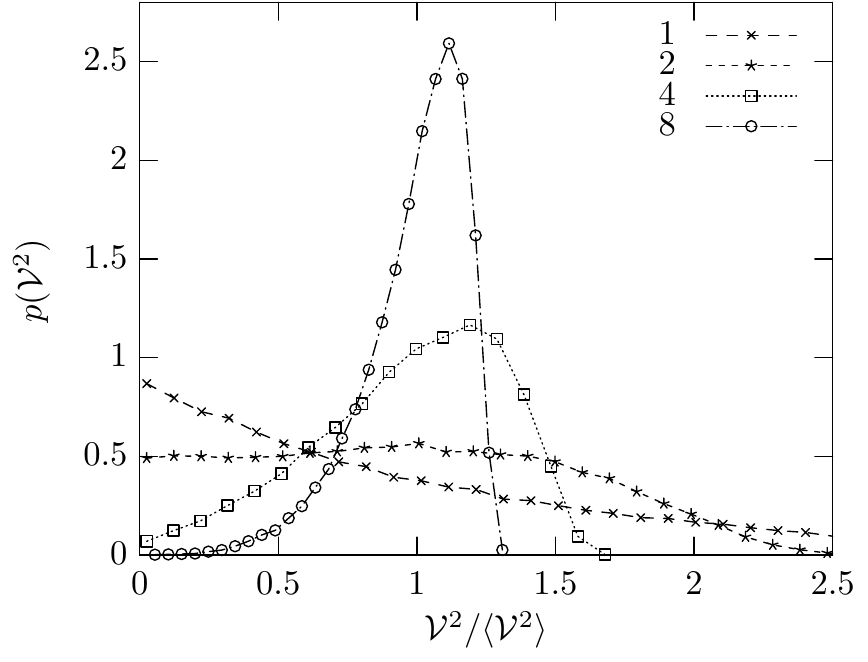}
	\caption{Probability distributions $p(\V^2)$ of the visibility as
	a function of 	$\V^2/\langle\V^2\rangle$ for $T_c/T=
	1,2,4,8$, for an anisotropy $a=1/10$. Unlike the distribution in
	the isotropic case, the distribution for a strongly anisotropic
	sample undergoes strong shape modifications as $T_c/T$ increases.}
	\label{fig:Kcrossaniso}
\end{figure}

The evolution of the probability distribution with decreasing temperature in
the strongly anisotropic case is shown in Fig.\,\ref{fig:Kcrossaniso} (for
$a=1/10$). Unlike the isotropic case, the value $\V^2=0$ is actually quite
probable at temperatures close to the critical point so that the shape of the
distribution shows a clear qualitative change as the temperature is lowered
towards the asymptotic regime. This observation is in qualitative agreement
with preliminary experimentally data taken at ENS \cite{hadzunpu}. Note that
as long as the probability for a vanishing visibility stays finite, the shape
of the distribution depends on the expectation and is thus explicitly
cutoff-dependent. However, since this dependence is only logarithmic we expect
the qualitative evolution of the shape to be insensitive to the precise
value of the cutoff.

\begin{figure}[htbp]
	\centering
	\includegraphics[width=\linewidth]{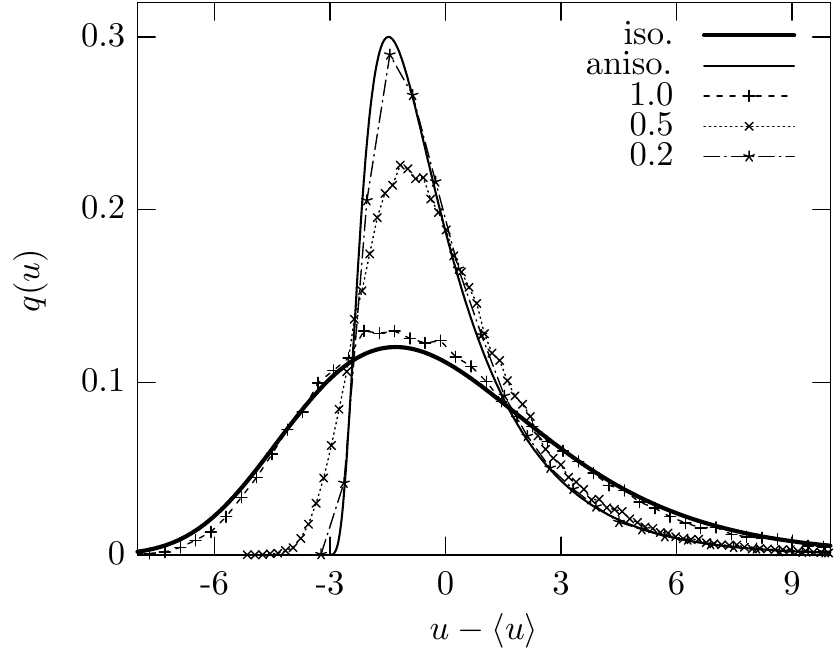}
	\caption{Probability density $q(u)$ for
	$a=1,1/2,1/5$ as well as the
	asymptotic distributions for the isotropic (heavy line) and strongly
	anisotropic (thin line) regime. For $a\leq 1/10$ (not shown), the
	results become indistinguishable from the strongly anisotropic limiting
	case. A value of $T_c/T=256$ has been
	chosen to ensure that the distributions are well within the asymptotic
	low temperature regime.}
	\label{fig:crossisoaniso}
\end{figure}

In turn, Fig.\,\ref{fig:crossisoaniso} shows the evolution of the probability
distribution in the asymptotic low temperature regime ($T=T_c/256$) with 
changing aspect ratio, confirming our earlier statement that the strongly
anisotropic regime is reached already for $a\lesssim 1/5$.

\subsection{Harmonically trapped sample}
\label{sec:ho}

As we have pointed out in the preceding discussion, the
cumulants (and hence the shape of the distribution) are dominated by the
excitations with the lowest frequency. The periodic boundary conditions we 
have used up to this point are thus quite artificial: even for a homogeneous system,
going over to different boundary conditions will have 
effects on the shape of the distribution (see \cite{imam2008} for examples).
However, the differences are mere numerical factors appearing in the cumulants
(for an example, the variance in a 3d homogeneous sample is smaller by a factor
of 1.67 when Dirichlet boundary conditions are used instead of periodic ones
\cite{zwer2004}), the dependence on physical parameters remains unaltered.

We now consider the geometry most relevant for actual experiments: a sample
which is harmonically trapped. For simplicity, we consider an isotropic trap
with trapping frequency $\omega$.  If the particle number is large enough to
warrant $N\tilde{g}\gg 1$ \cite{bloc2008} (which is readily fulfilled for
$\tilde{g}\sim 0.1$ and $N\sim 10^3$) and the temperature is sufficiently low,
the density distribution takes the form of a Thomas-Fermi profile
\begin{equation}
	\bar{n}(r)=\bar{n}(0)\left[1-\left(\frac{r}{R}\right)^2\right]
	\theta(R-r)
	\quad ; \quad
	R=\sqrt{\frac{2g\bar{n}(0)}{m\omega^2}} \ .
	\label{eq:TFprofile}
\end{equation}
In the following, we will assume the entire sample is superfluid and we need
not distinguish between the superfluid and the total density. This requires the
temperature to be substantially below $T_{\text{BKT}}= (4\pi
N\omega^2/\tilde{g}D_c^2)^{1/2}$, where the phase space density 
$D=\bar{n}\lambda_T^2$ reaches the critical value $D_c=\log(380/\tilde{g})$
of the BKT transition in the trap center \cite{prok2001b}. Moreover, we want
to be in the regime of near-unity visibility so that we must fulfill
$\delta\phi(R) =(\bar{n}(0)\lambda_T^2)^{-1}\log(R^2/\xi^2)\ll 1$
\cite{bloc2008}, or equivalently,
\begin{equation}
	T\ll T_\phi = \frac{1}{\log\left(2\tilde{g} N/\pi \right)}
	\sqrt{\frac{4\pi N\omega^2}{\tilde{g}}}\ .
	\label{eq:tphi}
\end{equation}
For $N=5000$, $\tilde{g}=0.1$ and $\omega=2\pi\times 20\,\text{Hz}$, one
obtains $T_{\text{BKT}}=95\,\text{nK}$ and $T_\phi=132\,\text{nK}$ (for lower
interaction constants, both are even higher) so that the low-temperature regime
we are considering is within experimental reach.

In analogy to the calculation of the low-energy modes in 3d \cite{stri1996},
the solutions $\psi_{n,l}(r,\theta)$ of the Euler-Lagrange equation
\eqref{eq:eulerlagrange} with open boundary conditions yield the 
frequencies
\begin{equation}
	\omega_{n,l}=\omega\sqrt{\frac{n(n+2)-l^2}{2}}\ .
	\label{eq:omeganl}
\end{equation}
Here, $n=0,1,2\dots$ is a radial and $l=-n,-n+2,\dots,n-2,n$ an azimuthal
index \cite{ho1999}. The eigenmodes are of the general form
\begin{equation}
	\psi_{n,l}(r,\theta)=\frac{P_{n,|l|}(r/R)}{R}\times
	\begin{cases}
		\cos(l\theta)/\sqrt{\pi} & l > 0\\
		1/\sqrt{2\pi} & l = 0 \\
		\sin(|l|\theta)/\sqrt{\pi} & l < 0
	\end{cases}\ ,
	\label{eq:psinl}
\end{equation}
where $P_{n,|l|}(x)=\sum_{k}a^{(n,|l|)}_k x^k$ are polynomials the coefficients
of which may be obtained from the recursion relation
\begin{equation}
	a^{(n,|l|)}_{k+2}[l^2-(k+2)^2]=a^{(n,|l|)}_k[n(n+2)-k(k+2)]\ .
	\label{eq:recursion}
\end{equation}
Hence, the $P_{n,|l|}(x)$ are either even or odd, and the highest and lowest
occurring power of $x$ are $n$ and $|l|$, respectively. The magnitude of the
lowest coefficient is fixed by the normalization condition
$\int_0^1\d x\,xP_{n,|l|}(x)^2 = 1$.

The eigenmodes satisfy the orthogonality relations
\begin{equation}
	\int_0^{2\pi}\d\theta\int_0^R\d r\, r\psi_{n,l}(r,\theta)
	\psi_{n',l'}(r,\theta)=\delta_{n,n'}\delta_{l,l'}
	\label{eq:orth}
\end{equation}
on a disk of radius $R$. For different $l$, the orthogonality is assured by the
azimuthal part, for equal $l$, by the radial part of the eigenfunctions.

A particularly important subset of the eigenfunctions is formed by the
\emph{surface modes}
\begin{equation}
	\psi_{n,\pm n}(r) = \sqrt{2(n+1)}\frac{r^n}{R^{n+1}}
	\begin{cases}
		\cos(n\theta)/\sqrt{\pi} & l = n\\
		\sin(n\theta)/\sqrt{\pi} & l = -n
	\end{cases}
	\label{eq:surfacemodes}
\end{equation}
with eigenfrequencies $\omega_{n,\pm n}=\sqrt{n}\omega$. By inspecting equation
\eqref{eq:omeganl}, one finds that most of the lowest-lying modes are such
surface modes. This is physically intuitive, since the phase stiffness as given
by equation \eqref{eq:TFprofile} is lower close to the rim so that low-energy
excitations should live on the boundary of the sample. It is thus to be 
expected
that the condensate fraction and hence the interference contrast is dominated 
by the behavior of the surface modes.

If we substitute the eigenmodes and eigenfrequencies of the surface modes into
the definitions of $\epsilon$ and $u$, we obtain (noting that 
$\psi_{n,l}(\vec{r})^2/\omega_{n,l}^2$ takes its maximum for $|l|=1$ and $r=R$)
\begin{equation}
	\epsilon^2 = \frac{4}{\pi}\frac{mT}{\rhos(0)}
	\quad ; \quad
	u=\pi\frac{\rhos(0)}{mT}(1-\V^2)\ .
	\label{eq:epsho}
\end{equation}
Once again, as in the homogeneous case, there is no explicit dependence on the
interaction constant. However, in the present case, there is an \emph{implicit}
dependence on the interaction constant since the latter defines the geometry of
the sample and the density in the center is given by $\bar{n}(0)=
mN\omega^2/\pi g$.  

Carrying out the expansion \eqref{eq:firstorder}, we find that in the case of
harmonic trapping the integrals \eqref{eq:integrals} become nontrivial. One
readily finds that $I_{n,l,n',l'}=\delta_{l,l'}I_{n,n'}(|l|)$ and 
$J_{n,l}=\delta_{l,0}J_n$ which follows immediately from the azimuthal part of
the eigenfunctions. This shows already that there is no $I$ that ``couples''
different surface modes (they all differ in $l$) and no $J$ that involves any
surface modes. The diagonal elements for the surface modes are readily found to
be $I_{n,\pm n}(n)=2/(n+2)$. On closer inspection, one finds that the matrices
$I_{n,n'}(|l|)$ are tridiagonal in the sense that they are non-vanishing only
for $n'=n,n\pm 2$ (we recall that $n$ and $l$ are either both even or both
odd), likewise only $J_2=1/\sqrt{3}$ is finite.

Taken together with the physical intuition that the statistics should be
dominated by the surface modes, these results suggest that it should be a
reasonable approximation to disregard $J_2$ and the non-diagonal elements of
the matrices $I_{n,n'}(|l|)$ which would permit to carry out the integration
leading to equation \eqref{eq:charfunction}. In order to see whether this 
leads to correct results we evaluate the variance of $u$ first exactly, then in
the ``diagonal'' approximation. A straightforward calculation gives
\begin{equation}
	\langle u^2\rangle_c
	=2{\sum_{n,n',l}}'\frac{I_{n,n'}(|l|)^2}
	{\tilde{\omega}_{n,l}^2\tilde{\omega}_{n',l}^2}
	-\frac{1}{6}{\sum_n}'\frac{I_{2,n}(0)}{\tilde{\omega}_{n,0}}
	+\frac{1}{72}\ ,
	\label{eq:varuexact}
\end{equation}
which is reminiscent of a structurally similar expression for the variance
of condensate fluctuations in a 3d harmonic trap by Giorgini \etal
\cite{gior1998}, but with two additional terms which come from the finiteness
of $J_2$. Once again, the primes on the sums are reminders that the sums go
only over permitted values of the indices.

\begin{table}
	\centering
	\begin{tabular}{c|c|c}
		Approximation & $\langle u^2\rangle_c$ 
			& $\sqrt{\langle u^2\rangle_c}$ \\
			\hline\hline
		exact & 2.813 & 1.677 \\
		diagonal & 2.496 & 1.580 \\
		surface & 2.160 & 1.470
	\end{tabular}
	\caption{Variance and standard deviation of $u$ in the harmonically
	trapped case, exact and in two different 
	approximations.	\label{tab:varu}}
\end{table}

In table~\ref{tab:varu}, we give the numerical value of equation 
\eqref{eq:varuexact} calculated in three different manners: first exactly,
\ie, without
approximations apart from the numeric calculation, and in two different
approximations. In the ``diagonal'' approximation, we disregard the two last
terms which are generated by $J_2$ and take into account only the diagonal
elements of $I_{n,n'}(|l|)$ in the first term. The ``surface'' approximation
goes even one step further by dropping all contributions except those coming
from the surface modes.
We see that the diagonal approximation is quite satisfactory (we recall that the
higher cumulants are increasingly dominated by the lowest-lying modes so that
we expect the approximation to improve with increasing cumulant order) and
even the elementary surface approximation fares reasonably well.

In the diagonal approximation, $u$ reads
\begin{equation}
	u={\sum_{n,l}}'\frac{t_{n,l}^2}{\tilde{\omega}_{n,l}^2}I_{n,n}(|l|)
	\label{eq:udiag}
\end{equation}
and the spectral zeta function is
\begin{equation}
	\zeta_{\{n,l\}}(s)={\sum_{n,l}}'\left( \frac{I_{n,n}(|l|)}
	{\tilde{\omega}_{n,l}^2} \right)^s\ .
	\label{eq:zetadiag}
\end{equation}
In the surface approximation, this can be written explicitly as
\begin{equation}
	\zeta_{\{n\}}^{(\text{surf})}(s)=2\sum_{n=1}^\infty
	\left( \frac{2}{n(n+2)} \right)^s
	\approx 2\left( \frac{2}{3} \right)^s\zeta(3s/2)\ .
	\label{eq:surfzeta}
\end{equation}
The approximation on the right-hand side is better than within one percent
for all $s\geq 2$. The global factor of $2$ comes from the two-fold
degeneracy of the surface modes. As in the strongly anisotropic rectangle
case, it compensates the global factor of $1/2$ in equation 
\eqref{eq:defcumgen}.

While it does not seem possible to derive a closed-form expression for the
associated probability distribution, the form of the spectral zeta function
suggests that the distribution should be something intermediate between a
Gumbel distribution [where the spectral zeta function is proportional to
$\zeta(s)$] and a Jacobi distribution [where the spectral zeta function is
proportional to $\zeta(2s)$], \ie, more asymmetric than the former, but less
asymmetric than the latter. Of course, the contributions from the neglected
modes will render the actual distribution somewhat more symmetric than this
argument suggests, but it should remain qualitatively valid.
\begin{figure}[htbp]
	\centering
	\includegraphics[width=\linewidth]{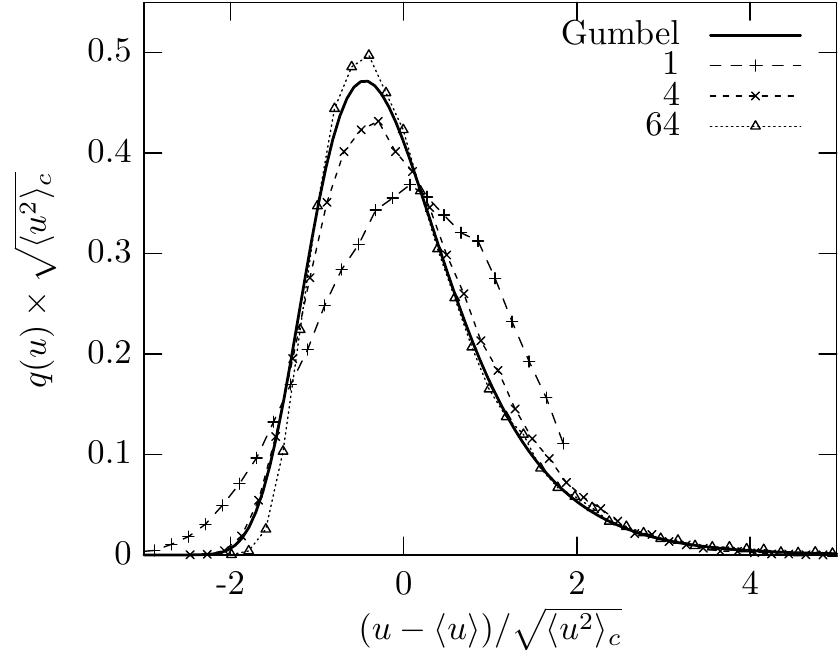}
	\caption{Scaled distribution $q(u)$ for the harmonically trapped
	case, for $T_c/T=1,4,64$. A Gumbel distribution (not a convolution of 
	Gumbel 
 	functions) scaled to have unit variance is shown for comparison
	(continuous line). The sharp fall-off at the right end of
	the $T=T_c$ curve corresponds to zero visibility while the
	$T=T_c/64$ curve is already well within
	the asymptotic low-temperature regime. Quite amusingly, the profile
	for $T=T_c/T=16$ (not shown) comes out almost superposed with a Gumbel 
	distribution while the asymptotic distribution comes quite close to 
	it, but remains more sharply peaked, in agreement with the arguments
	presented in the text.}
	\label{fig:uscaledHO}
\end{figure}
This is supported by numerical results which are calculated without 
approximations, as can be seen in figure~\ref{fig:uscaledHO}.

\section{Distribution at the critical point}
\label{sec:critpoint}

A quite interesting aspect associated with the statistics of interference
amplitudes is the possibility to measure the universal probability distribution
of the order parameter near a critical point. Indeed, as has been shown in
section \ref{sec:intstats},  the visibility is identical with the condensate
fraction provided the integration length is much larger than the interparticle
spacing. Our calculation of the resulting visibility distribution in the
previous section is valid deep on the Bose-condensed regime, where the
visibility is close to one. 

\begin{figure}[htbp]
	\centering
		\includegraphics[width=.8\linewidth]{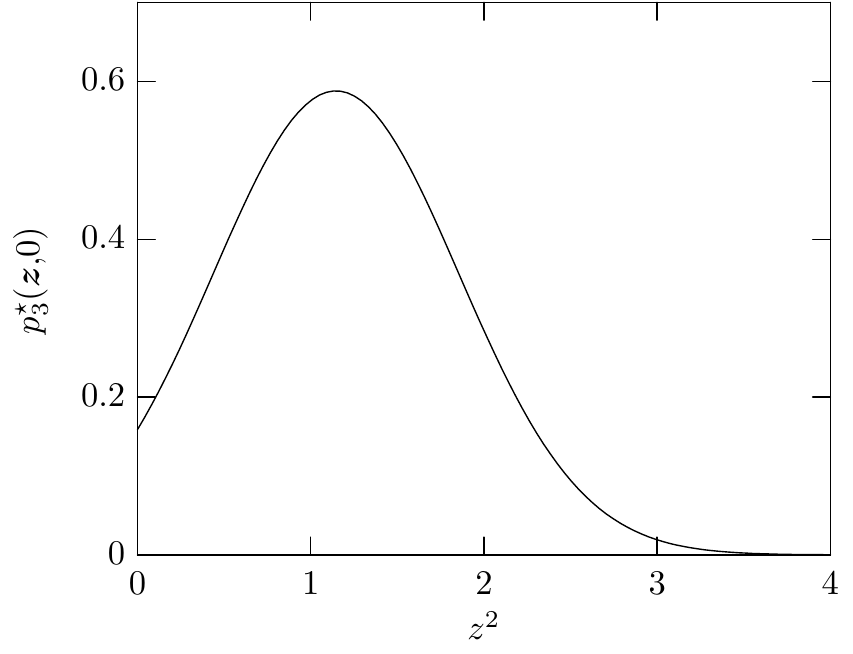}
		\caption{The universal distribution $p^\star_3(\vec{z},0)$
		for the order parameter $\vec{z}$ in 3d directly at the
		critical point as calculated in \cite{chen1996}.}
	\label{fig:dohm}
\end{figure}

In the following, we want to discuss the situation close to the 
critical point of Bose-Einstein-condensation, that is in a regime 
where the system size $L$ is large compared to microscopic lengths
but of the same order or smaller than
the correlation length $\xi$ of the infinite system. Mathematically, 
this may be expressed by a dimensionless parameter 
\begin{equation}
	x=tL^{1/\nu}=\pm\left(\frac{L}{\xi}\right)^{1/\nu}=\mathcal{O}(1)
	\label{eq:critical}
\end{equation} 
that measures the deviation from the critical point due to the finite system
size.  Here $t=(T-T_c)/T_c$ is the dimensionless distance from the bulk
critical temperature $T_c$ and $\nu\approx 0.672$ the critical exponent that
characterizes the divergence $\xi\sim |t|^{-\nu}$ of the correlation length in
a 3d BEC \cite{camp2001,buro2006}. This exponent has in fact been measured also
in dilute ultracold gases \cite{donn2007}. Quite generally, finite size scaling
predicts that the probability distribution of a two-component order parameter
$\vec{s}$ (for a BEC, $\vec{s}^2=n_0$ is the condensate fraction) in the
critical regime has a scaling form \cite{bind1981,brez1985} 
\begin{equation}
	p(\vec{s}, t, L)=L^{2y}p_d^{\star}\left(\vec{s}L^y, tL^{1/\nu}\right)
	\ .
	\label{eq:scaling}
\end{equation} 
Here,
\begin{equation}
	y(d)=(d-2+\eta)/2
	\label{eq:yd}
\end{equation} 
is related to the standard anomalous dimension $\eta$ of the XY-model while
$p_d^{\star}(\vec{z},x)$ is a universal, non-Gaussian distribution that only
depends on $\vec{z}^2=n_0L^{2y}$.  The existence of such a distribution for
properly scaled block-spin variables $\vec{s}L^y$ with finite moments of
arbitrary order, is in fact a basic assumption of the renormalization group
approach to critical points, as emphasized by Parisi \cite{pari1987}. Precisely
at the critical point, where $x=0$, the distribution is determined by the
effective potential $V_{\text{eff}}^{\star}(\vec{z})$ of the underlying field
theory at the fix-point via 
\begin{equation}
	p_d^{\star}(\vec{z},x=0)\sim
	\exp\left[-V_{\text{eff}}^{\star}(\vec{z})\right]
	\label{eq:pstar}
\end{equation} 
Within the standard two-component $\Phi^4$-theory, the distribution
$p_3^{\star}(\vec{z},x)$ in the 3d case has been calculated by Chen \etal
\cite{chen1996}, taking into account the singularities associated with the
Goldstone mode.  The resulting distribution at the critical point $x=0$ is
shown in Fig.~\ref{fig:dohm}. It exhibits a maximum at
$\vec{z}^2\simeq 1.146$.  The most probable value for the number of particles
in the condensate therefore scales as $\bar{N}_0\simeq L^{2-\eta}$ which is
sub-extensive, as expected right at $T_c$. The average typical visibility 
\begin{equation}
        \sqrt{\langle\V^2\rangle}\sim 1/L^{1+\eta}
	\label{eq:visibilityaverage}
\end{equation} 
at the critical point will therefore vanish with an anomalous power of the
integration length $L$, \ie,  basically like $1/L$ because $\eta\simeq 0.03$ is
rather small.  Moreover, the fact that the variance of the variable $\vec{z}^2$
is a universal constant of order one, the so-called Binder cumulant
\cite{bind1981}, implies that the fluctuations of the visibility scale with the
same anomalous power of the integration length as the average.

In the 2d case, there is no breaking of a continuous symmetry at finite
temperature. Instead, there is a Berezinskii-Kosterlitz-Thouless transition to
quasi long range order below $T_c$, where the gas is a proper superfluid but
not a BEC. The absence of a finite correlation length below $T_c$ in this case
does not allow to define a simple analog of the variable $x$ in
\eqref{eq:critical}. Right at the BKT-transition, however, one expects again a
universal order parameter distribution function $p_2^{\star}(\vec{z})$ for the
variable $\vec{z}^2=n_0L^{\eta}$, since $y=\eta/2$ in two dimensions.  In
contrast to the situation discussed in section \ref{sec:analyt}, where the 
distribution
of the interference contrast has been calculated  deep in the superfluid
regime and the visibility is close to one, the distribution
$p_2^{\star}(\vec{z})$ with its anomalous scaling applies to 2d Bose gases
whose size is much larger than the phase coherence length $\ell_{\phi}$. The
thermal phase fluctuations then imply an average condensate fraction $\langle
n_0\rangle\sim L^{-\eta}$ which decreases with system size.  The typical value
$\sqrt{\langle\V^2\rangle}\sim L^{-\eta}$ of the visibility is therefore close
to zero. In fact, since the 2d superfluid phase corresponds to a line of
critical points at \emph{any} $T<T_c$, this behavior of the average visibility
is valid at \emph{arbitrary} temperature below $T_c$ in the limit $L\rightarrow
\infty$ with a temperature dependent exponent $\eta(T)$ which reaches its 
critical value $\eta_c=1/4$ at $T_c$.
Note that, independent of the precise form of the distribution
$p_2^{\star}(\vec{z})$, the very existence of a scaling variable
$\vec{z}^2=n_0L^{\eta}$ immediately implies the sub-extensive scaling $\langle
N_0\rangle\sim L^{2-\eta}$ of the average number of particles in the
condensate for an interacting 2d Bose gas and its anomalous fluctuations  
Var$\, N_0\sim\langle N_0\rangle^2$ \cite{meie1999} in the thermodynamic limit. 

In order to observe this anomalous scaling, the system size must be large
compared to the phase coherence length, $L\gg\ell_\phi=\xi e^{1/2\eta(T)}$. At
the critical point, this is readily fulfilled for typical system sizes of the
order of some \unit{10}{\micron} and healing lengths of the order of
\unit{0.1}{\micron}.  By contrast, for $T\ll T_c$, the system size required to
be in the anomalous scaling regime rapidly exceeds experimentally feasible
values. Therefore one has $L\ll\ell_\phi$ in practice and the visibility
distribution can be determined by an expansion around $\V^2\approx 1$ as done
in section \ref{sec:analyt}.

\section{Conclusion}

In conclusion, we have shown that interference experiments may be used as a
direct measurement of the statistics of the condensate fraction in ultracold
Bose gases. Unlike in interference experiments in classical optics, where the
fringe visibility is determined by a deterministic cross correlation function
of the optical fields \cite{zern1938, wolf2007}, the interference contrast of
matter waves is a quantum observable.  Repeated experiments with identically
prepared condensates therefore produce a statistical distribution of values
instead of a reproducible single value.  The resulting distributions are
non-Gaussian even in the thermodynamic limit and have been calculated
explicitly for 2d Bose gases at temperatures such that their effective
condensate fraction is close to one.  Quite generally, the interference
contrast is a self-averaging observable in situations with long range phase
coherence.  Our findings for the 2d strongly anisotropic case are in
qualitative agreement with preliminary data taken at ENS \cite{hadzunpu}.
Clearly, a quantitative comparison between theory and experiment is needed to
verify our predictions. In particular, the interference statistics might be
used as a precise thermometer of the gases, similar to what has been achieved
in 1d gases \cite{hoff2008}.  A quite interesting open problem, both from a
theoretical and an experimental point of view, is the analysis of the
interference contrast near the  transition to the normal phase. It offers
the possibility to directly measure the distribution of the order
parameter\footnote{Note that for 2d gases there is no true order parameter, yet
there is a nontrivial distribution of the number of particles at zero
momentum.} near the critical point, a quantity that is very hard to measure
otherwise.  

\acknowledgements
The authors acknowledge helpful discussions with M.~Holzmann and B.~Spivak.  We
are very grateful to Z.~Hadzibabic, P.~Kr{\"u}ger and J.~Dalibard for providing
us with unpublished experimental data \cite{hadzunpu}. Part of this work has
been supported by the DFG research unit ``Strong Correlations in Multiflavor
Ultracold Quantum Gases''.

\appendix

\section{Universal scaling in 3d}
\label{app:scale3d}

As was shown in section \ref{sec:intstats}, the distribution of the condensate
number, which is related to the intensive  two-component vector order parameter
$\vec{s}$ that describes Bose-Einstein-condensation from the point of view of
statistical physics by $N_0=L^3\vec{s}^2$ is not a simple Gaussian. This is a
result of the fact that in the case where the broken symmetry is continuous,
the order parameter correlation length is infinite for all temperatures below
$T_c$ \cite{zwer2004}. The universal distribution function of the condensate
number below $T_c$ is in fact contained in the result \eqref{eq:defcumgen} for
the logarithm of the characteristic function of the random variable
$u=2(1-\V^2)/\epsilon^2$. In 3d, the small parameter $\epsilon$ defined in
equation \eqref{eq:epsclass} can be written in the form
\begin{equation}
	\epsilon^2 = \frac{1}{\pi^2}\frac{\xi_{\text{J}}(T)}{L}\ ,
	\label{eq:eps3d}
\end{equation}
where we have introduced the Josephson length $\xi_{\text{J}}=mT/\rhos$.  For
any finite temperature therefore, $\epsilon$ goes to zero for a system size $L$
much larger than the Josephson length. The universal distribution $p(u)$ for
the fluctuating variable
\begin{equation}
	u= \frac{2\pi^2 L}{\xi_{\text{J}}}\left(1-\frac{N_0^2}{N^2}\right)
	\label{eq:uN0}
\end{equation}
which determines the distribution of the condensate number of a 3d BEC below
$T_c$ is fixed by the exact cumulants given in \eqref{eq:cumulants}.  It
depends on the 3d spectral zeta function
\begin{equation}	
	\zeta_{\{\lambda\}}(s)={\sum_{l_1,l_2,l_3}}'
	\frac{1}{(l_1^2+l_2^2+l_3^2)^s}
	=\sum_{n=1}^\infty \frac{A_3(n)}{n^s}	\ ,
	\label{eq:pbczeta3}
\end{equation} 
for which, unfortunately, no closed-form expression seems to exist
\cite{zuck1974}.  It is evident, however, that $\zeta_{\{\lambda\}}(s)$ is
convergent for all $s>3/2$ and thus all cumulants except the first are finite.
The variable  with a proper, non-Gaussian distribution in the limit
$L\to\infty$ is thus $n_0^2\cdot L/\xi_J$ which implies that the condensate
fraction is a self-averaging variable. Its fluctuations, however, are not of
order $1/\Omega$ as usual but only decay like $1/L^2$ \cite{zwer2004}. Since
all higher cumulants including the variance are finite and have the same
scaling with system size $L$, the ratios $\langle u^s\rangle_c/\langle
u^2\rangle_c^{s/2}$ are constant and finite. A
special case of  this result has in fact been found by Kocharovsky \etal
\cite{koch2000}, who calculated the cumulants of the number of condensed atoms
in a 3d BEC within a Bogoliubov approach. In particular, to leading order in
$\epsilon$, our cumulants from equation \eqref{eq:cumalpha} agree with theirs,
showing the close connection between the condensed fraction and the
interference amplitude. 

\section{The 1d case at zero temperature}

In this appendix, we discuss the case of a homogeneous 1d Bose gas at vanishing
temperature. Using $c=\sqrt{g\rhos/m}$ and substituting $\rhos/m=cK/\pi$, where
$K$ is the dimensionless Luttinger parameter, the hydrodynamic action
\eqref{eq:action} governing the phase difference $\phi=\varphi_2-\varphi_1$
of two interfering 1d Bose gases has the form
\begin{equation}
	S_0[\phi]=\frac{K}{4\pi c}\int_0^L\d x\int_0^\beta\d\tau\,
	\left\{  
	\left[\partial_\tau\phi\right]^2
	+c^2\left[\partial_x\phi\right]^2
	\right\}\ ,
	\label{eq:Seff}
\end{equation}
where we have kept $\beta$ finite. Using the Fourier expansion
\begin{equation}
	\phi(x,\tau)=\frac{1}{\sqrt{\beta L}}\sum_k\sum_{n=-\infty}^\infty
	\phi_k(\omega_n)e^{i(kx-\omega_n\tau)}\ ,
	\label{eq:fourierseries}
\end{equation}
where $k=2\pi l/L$ with $l\in\mathbb{Z}$, and $\omega_n=2\pi n/\beta$ are the
bosonic Matsubara frequencies, the action takes the diagonal form
\begin{equation}
	S_0[\phi]=\frac{K}{4\pi c}\sum_{k,n}(c^2k^2+\omega_n^2)
	|\phi_k(\omega_n)|^2\, .
	\label{eq:diagS}
\end{equation}

The
generating function $p(\sigma)=\langle e^{i\sigma\V^2}\rangle$ for the square
of the visibility requires calculating a functional integral with a
perturbation 
\begin{equation}
	S_1=\frac{i\sigma}{N^2}\int\d x\int\d x'\,\bar{n}(x)\bar{n}(x')
	\cos\left[ \phi(x)-\phi(x') \right]
	\label{eq:perturb}
\end{equation}
to the action \eqref{eq:Seff}. This perturbation only contains the phase difference 
$\phi(x)\equiv \phi(x,0)$ on the boundary in imaginary time $\tau$.  Except for
$\phi(x)=\phi(x,\tau=0)$ all variables are therefore Gaussian and can be integrated out. 
The problem then is completely analogous to that of backscattering from a single 
impurity in a Luttinger liquid discussed by Kane and Fisher 
\cite{kane1992b}. 

Upon elimination of the modes $\phi_k(\tau\neq 0)$, one obtains the reduced
free action
\begin{equation}
	S_0[\phi]=\frac{K}{2\pi}\sum_{k=-\Lambda}^\Lambda
	|k| \left|\phi(k)\right|^2
	\label{eq:S0q}
\end{equation}
for the remaining, non-Gaussian degrees of freedom,
where we have explicitly written the ultraviolet cutoff $\Lambda$. This
corresponds to a \emph{non-local} action in space of the form
\begin{equation}
	S_0[\phi(x)]=\frac{\alpha}{8\pi^2}\int\d x\int\d x'
	\left( \frac{\phi(x)-\phi(x')}{x-x'} \right)^2
	\label{eq:S0x}
\end{equation}
that arises in $\tau$ space for dissipative quantum mechanics of a single
particle \cite{fish1985} or in the study of nontrivial ground states of 
open strings \cite{call1990}.
Equations \eqref{eq:S0q} and \eqref{eq:S0x} are
equivalent if the associated dimensionless strength $\alpha$ of the 
dissipation is related to the Luttinger parameter by $\alpha=2K$.

Following the arguments of section \ref{sec:analyt}, the distribution of the
interference contrast can be calculated analytically in the limit
$\epsilon^2=1/K\ll 1$ by expanding $S_1$ in equation
\eqref{eq:perturb} to second order in $\phi$. Again, it is then natural to
consider the characteristic function $q(\sigma)=\langle
e^{i\sigma(1-\V^2)}\rangle$ which corresponds to a perturbation (for a
homogeneous system with $\bar{n}(x) =N/L$)
\begin{equation}
	\hat{S}_1[\phi]=-\frac{i\sigma}{2L^2}\int\d x\int\d x'\,
	\left[\phi(x)-\phi(x')  \right]^2\ .
	\label{eq:newperturb}
\end{equation}
Substituting the Fourier series representation for $\phi(x)$, this becomes
\begin{equation}
	\hat{S}_1[\phi]=-\frac{i\sigma}{L}\sum_{k\neq 0}\left|\phi_k\right|^2
	\label{eq:S1q}\ .
\end{equation}
The functional integral is now Gaussian and can be evaluated exactly, giving
\begin{equation}
	q(\sigma)
%	=\prod_{q\neq 0}\left( 
%	\frac{K|q|/\pi-2i\sigma/L}{K|q|/\pi}\right)^{-1/2}
	=\prod_{k>0}\left( 1-\frac{2\pi i\sigma}{K kL} \right)^{-1}\ ,
	\label{eq:spectdet}
\end{equation}
or
\begin{equation}
	\log q(\sigma) = \frac{i\sigma}{K}\langle 1-\V^2\rangle
	+\sum_{s=2}^\infty\frac{1}{s}
	\left( \frac{i\sigma}{K}\right)^s\zeta(s)\ ,
	\label{eq:logqw1d}
\end{equation}
where again the expectation is explicitly cutoff-dependent. 
%Replacing the sum
%for $s=1$ by an integral from $1$ to $\Lambda L$, one readily finds that the
%condensate fraction goes like $1-1/K\log(\Lambda L)$ while 
Comparison with
equation \eqref{eq:pxgumbel} shows that the variable
$K(\langle\V^2\rangle-\V^2)$ has a Gumbel distribution of the normalized form
given in equation  \eqref{eq:pxgumbel} as derived in \cite{imam2006}. 

Note that the action $S_0+S_1$ as given by equations \eqref{eq:Seff} and
\eqref{eq:perturb} differs from the action of the boundary sine-Gordon model
that appears for dissipative quantum mechanics in a (purely imaginary) periodic
potential. Instead it corresponds to a classical 1d XY-model model with
infinite range interactions. However, a mapping to a sine-Gordon model (relying
on a Hankel transform rather than a Fourier transform of the probability
distribution) is possible in the thermodynamic limit and has been used by 
Gritsev \etal in \cite{grit2006} to calculate the distribution function of the
interference contrast for \emph{arbitrary} values of the Luttinger parameter
$K$.

\end{document}